\documentstyle[jkas2]{article}

\runningauthor{KANG AND JONES}
\runningtitle{COSMIC-RAY ACCELERATION} 
\beginpage{159}
\endpage{174}

\def\etal{{\it et al.~}}
\def\eg{{\it e.g.,~}}
\def\ie{{\it i.e.,~}}

\def\cm3{~{\rm cm^{-3}}}

\def\lsim{\mathrel{  
        \raise0.3ex\hbox{$<$}\kern-0.75em{\lower0.65ex\hbox{$\sim$}}}}
\def\gsim{\mathrel{
        \raise0.3ex\hbox{$>$}\kern-0.75em{\lower0.65ex\hbox{$\sim$}}}}

\begin{document}

\title{Acceleration of Cosmic Rays at Large Scale Cosmic Shocks 
in the Universe}

\author{HYESUNG KANG$^1$ and T. W. JONES$^2$}

\address{$^1$Department of Earth Sciences, Pusan National University, Pusan  609-735, Korea \\
$^2$Department of Astronomy, University of Minnesota, Minneapolis,
      MN 55455, USA \\
{\it E-mail: kang@uju.es.pusan.ac.kr,
and twj@msi.umn.edu }}
\address{\normalsize{\it (Received Oct 18, 2002; Accepted Oct. 25, 2002)}}

\abstract{
Cosmological hydrodynamic simulations of large scale structure in the
universe have shown that accretion shocks and merger shocks form due to
flow motions associated with the gravitational collapse of nonlinear
structures.  Estimated speed and curvature radius of these shocks could 
be as large as a few 1000 km/s and  several Mpc, respectively. 
According to the diffusive shock acceleration theory, populations of cosmic-ray
particles can be injected and accelerated to very high energy by 
astrophysical shocks in tenuous plasmas.
In order to explore the cosmic ray acceleration at the cosmic shocks,
we have performed nonlinear numerical simulations of cosmic ray (CR) modified shocks
with the newly developed CRASH (Cosmic Ray Amr SHock) numerical code.
We adopted the Bohm diffusion model for CRs, based on the hypothesis that strong
Alfv\'en waves are self-generated by streaming CRs.
The shock formation simulation includes a plasma-physics-based
``injection'' model that transfers a small proportion of the thermal proton flux
through the shock into low energy CRs for acceleration there.
We found that, for strong accretion shocks, CRs can absorb most of shock 
kinetic energy and the accretion shock speed is reduced up to 20 \%, compared
to pure gas dynamic shocks. 
For merger shocks with small Mach numbers, however, the energy transfer to
CRs is only about 10-20 \% with an associated  CR particle fraction of $10^{-3}$.
Nonlinear feedback due to the CR pressure is insignificant in the latter
shocks. 
Although detailed results depend on models for the particle diffusion and 
injection, these calculations show that cosmic shocks 
in large scale structure could provide acceleration sites of extragalactic
cosmic rays of the highest energy. 
}

\maketitle

\section{INTRODUCTION}

Shocks are ubiquitous in astrophysical environments: a few examples are
Earth's bow shock, interplanetary shocks, stellar wind terminal shocks, 
supernova remnants, shocks in radio jets, merger shocks in intracluster
media, and accretion shocks associated with large scale structure 
formation. 
Most astrophysical shocks are so-called ``collisionless shocks'' which form 
in a tenuous plasma via electromagnetic ``viscosities,'' 
\ie collective electromagnetic interactions
between the particles and the fields.
Hence the magnetic field, especially its irregular component, is
vital to the shock formation process.
Our discussion will focus on a quasi-parallel shock, in which the direction of 
propagation is almost parallel to the magnetic field lines.
According to plasma simulations of quasi-parallel shocks (Quest 1988), 
the particle 
velocity distribution has some residual anisotropy in the local fluid frame 
due to the incomplete isotropization during the collisionless shock formation 
process and so some particles can stream back upstream of the shock. 
Streaming motions of high energy particles against the background fluid generate 
strong MHD Alfv\'en waves upstream of the shock, which
in turn scatter particles and prevent them from escaping upstream 
(e.g., Wentzel 1974; Bell 1978;Quest 1988;Lucek \& Bell 2000).
Due to these self-generated MHD waves thermal particles are 
confined and advected downstream, while some suprathermal particles 
in the high energy tail of the Maxwellian 
velocity distribution may re-cross the shock upstream.
Then these particles are scattered back downstream 
by those same waves and can be accelerated further to higher energies 
via Fermi first order process.
Hence the nonthermal, cosmic-ray particles are natural byproducts of the
collisionless shock formation process and they are extracted from the 
shock-heated thermal particle distribution (Malkov \& V\"olk 1998,
Malkov \& Drury 2001).
This ``thermal leakage'' injection process has been observed well in
the Earth's bow shock and interplanetary shocks 
(Ellison, M\"obius, \& Paschmann 1990, Baring \etal 1997).
Also there have been observational evidence and theoretical studies
that strongly support the idea that the cosmic rays 
are accelerated via the ``diffusive shock acceleration (DSA)''
process at various astrophysical shocks
(\eg Drury 1983; Blandford \& Eichler 1987).

According to hydrodynamic simulations of large scale structure formation 
(\eg Kang \etal 1994a; Miniati \etal 2000),
accretion shocks are formed in the baryonic component around non-linear
structures collapsed from the primordial density inhomogeneities as a
result of gravitational instability.  Those structures can be identified as
pancake-like supergalactic planes, still denser filaments, and clusters of
galaxies that form at intersections of pancakes in any variants of the
many cosmological models.  These structures are surrounded by the hot gas heated by the
accretion shocks and the CRs (ions and electrons)
can be accelerated to very high energies
at these shocks via the first order Fermi DSA process (Kang, Ryu \& Jones 1996,
Miniati \etal 2001a, b).
The accretion shocks around the clusters of galaxies could involve
flows as fast as 
a few 1000 $km s^{-1}$ and, so, could be fast acceleration sites for the 
high energy cosmic rays up to several $\times 10^{19}$ eV, 
provided that the magnetic field around the clusters is order of microgauss. 
Norman, Melrose \& Achterberg (1995) also suggested that cosmic accretion
and merger shocks can be good acceleration sites for ultra-hight energy CRs 
above $10^{18.5}$ eV if a primordial field of order $\gsim 10^{-9} G$ exists,
or if microgauss fields can be self-generated in shocks.
In fact a magnetic field with a typical strength of a few microgauss and a
principal length scale of 10-100 kpc was detected in several clusters of
galaxies by the Faraday rotation of radiation from distant radio
sources (\eg  Kim, Kronberg, Tribble 1991, Kronberg 1994,
Taylor \etal 1994, Feretti \etal 1995, Clarke \etal 2001, Carilli \& Taylor 2002).
On the other hand, the magnetic fields derived from observed hard X-ray and 
EUV emissions, on the assumption that these are inverse Compton (IC) 
scattering of CMB photons, are
somewhat lower ($\sim 0.4$ microgauss) (Fusco-Femiano \etal 1999, 2000).
The origin and evolution of cosmic magnetic fields is beyond the scope of
this paper.
Magnetic fields may have been injected into the ICM by radio galaxies.
They may have been seeded at shocks in the course
of structure formation, and then stretched and amplified 
up to 0.1-1 microgauss levels by turbulent flow motions in ICM and 
also in filaments and sheets (Kulsrud \etal 1997, Ryu, Kang \& Biermann 1998). 

Recent observations in EUV and hard X-ray have revealed that some clusters possess
excess radiation compared to what is expected from the hot, thermal X-ray 
emitting ICM (\eg Sarazin \& Lieu 1998; Lieu \etal 1999; Ensslin \etal 1999;
Fusco-Femiano \etal 1999; Sarazin 1999).
One mechanism proposed for the origin of this component is the
IC scattering of cosmic microwave background photons
by CR electrons accelerated by merger shocks and accretion 
shocks around the clusters. 
Also it has been suggested that the diffuse gamma-ray background radiation
could originate from the same process
(Loeb \& Waxman 2000, Miniati 2002, Scharf \& Mukherjee 2002). 
The same mechanisms that are capable of producing CR electrons
may have produced CR protons, although the existence of CR protons
in the ICM has not yet been directly observed. 
The existing evidence for substantial CR populations in these
environments argues that nonthermal activities in the
ICM could be important in understanding the dynamical status and the evolution of
clusters of galaxies (Sarazin \& Lieu 1998; Lieu \etal 1999).
CR protons and electrons may provide a significant
pressure to the ICM, perhaps, comparable to the thermal gas pressure 
(Lieu \etal 1999, Colafrancesco 1999), 
as it is for the galactic CRs in the ISM of our galaxy.
Collisions of CR protons in the ICM generate a flux of $\gamma$-ray
photons through the production and subsequent decay of neutral pions.
While such $\gamma$-rays have not yet been detected from clusters
recent estimates have shown that $\gamma$-ray
fluxes from the nearest rich clusters, such as Coma, are within the
range of what may be detected by the next generation of $\gamma$-ray
observatories (Ensslin \etal 1997, Sreekumar \etal 1996, Miniati \etal 2001).

According to DSA theory a significant fraction (up to 90\%) of the kinetic 
energy of the bulk flow associated with the strong shock can be converted 
into CR protons, depending the CR injection rate
(Drury 1983, Jones \& Kang 1990, Berezhko, Ksenofontov, \& Yelshi 1995). 
If as much as $10^{-4} - 10^{-3}$ of the particle flux passing through the shock were
injected into the CR population, the CR pressure would dominate 
and the nonlinear feedback to the underlying flow would become substantial.
Recently Gieseler, Jones \& Kang (2000) have developed a novel numerical
scheme that self-consistently incorporates the thermal leakage injection
based on the analytic, nonlinear calculations of Malkov (1998).
This injection scheme, which has only one tightly restricted
adjustable parameters, has been implemented into the combined gas dynamics 
and the CR diffusion-convection code with the Adaptive Mesh Refinement 
technique by Kang, Jones \& Giesler (2002).
The CR injection and acceleration efficiencies at quasi-parallel, 
plane-parallel shocks were calculated with their new numerical code named
as CRASH (Comic-Ray Amr SHock) code.
They found that about $10^{-3}$ of incoming thermal particles are injected
into the CRs, that up to 60 \% of initial shock kinetic energy is transferred 
to CRs for strong shocks, and that the shock speed is reduced up to $\sim$ 17 \%
for shocks with Mach number greater 30.
These results have confirmed the findings of previous studies which adopted
a simpler injection model that included a fully adjustable free parameter 
(\eg Berezhko \etal 1995, Kang \& Jones 1995).

In this contribution we will present the numerical simulation results for
quasi-parallel shocks in 1D plane-parallel geometry with the physical
parameters relevant for the cosmological shocks emerging in large scale
structure formation of the Universe.
In the next section the details of numerical simulations, including
the basic equations, numerical method, and model parameters, will be
given. The simulation results are presented and discussed in \S III,
followed by a brief summary in \S IV.

\section{Numerical Methods}
\subsection{Basic Equations}
We solve the standard gasdynamic equations with CR pressure terms
added in the conservative, Eulerian
formulation for one dimensional plane-parallel geometry:

\begin{equation}
{\partial \rho \over \partial t}  +  {\partial (u \rho) \over \partial x} = 0,
\label{masscon}
\end{equation}

\begin{equation}
{\partial (\rho u) \over \partial t}  +  {\partial (\rho u^2 + P_g + P_c) \over
\partial x} = 0,
\label{mocon}
\end{equation}

\begin{equation}
{\partial (\rho e_g) \over \partial t}  +  {\partial (\rho e_g u + P_g u + P_c u
) \over \partial x} = - L(x,t),
\label{econ}
\end{equation}

\begin{equation}
{\partial S\over \partial t}  +  {\partial (uS) \over \partial x} =0,
\label{scon}
\end{equation}
where $P_{\rm g}$ and $P_{\rm c}$ are the gas and the CR pressure,
respectively, $e_{\rm g} = {P_{\rm g}}/[{\rho}(\gamma_{\rm g}-1)]+ u^2/2$
is the total energy density of the gas per unit mass and the rest of the
variables have their usual meanings.
The injection energy loss term, $L(x,t)$, accounts for the
energy of the suprathermal particles injected to the CR component at
the subshock.
Here, $S= P_g / \rho^{\gamma_g-1}$, the ``Modified Entropy'',
is introduced in order to follow the preshock adiabatic compression accurately 
in strong CR modified shocks.
We note that the equation (4) is valid only outside the dissipative subshock; 
\ie where the gas entropy is conserved.
Hence the modified entropy equation is solved outside the subshock, while
the total energy equation is applied across the subshock.

The diffusion-convection equation for
the pitch angle averaged CR distribution function, $f(p,x,t)$, {\eg Skilling 1975)
is given by
\begin{equation}
{\partial f\over \partial t}  + u {\partial f \over \partial x}
= {1\over3} ({\partial u \over \partial x})  p {\partial f\over
\partial p} +  {\partial \over \partial x} (\kappa(x,p)  {\partial f
\over \partial x}).
\label{diffcon}
\end{equation}
and $\kappa(x,p)$ is the spatial diffusion coefficient.
For convenience we always express the particle momentum, $p$ in
units $m_{\rm p}c$.
As in our previous studies, the function $g(p)=p^4f(p)$ is evolved instead
of $f(p)$ and $y = ln(p)$ is used instead of the momentum variable, $p$
for that step. 
Then the CR pressure is calculated from the nonthermal particle distribution
as follows:
\begin{equation}
P_{\rm c}= {4 \over 3} \pi m_{\rm p}c^2
          \int_{0}^{\infty} g(p) {{{\rm d} p}\over {\sqrt{p^2+1}}}.
\label{cr_pressure}
\end{equation}

\subsection{CRASH: CR/AMR Hydrodynamics Code}
Unlike ordinary gas shocks, the CR shock includes a wide range of length
scales associated not only with the dissipation into ``thermal plasma'',
but also with the nonthermal particle diffusion process.
Those are characterized by the so-called diffusion lengths,
\begin{equation}
D_{\rm diff}(p) = \kappa(p)/u,
\end{equation}
where $\kappa(p)$ is the spatial diffusion coefficient for CRs of momentum $p$,
and $u$ is the characteristic flow velocity (Kang \& Jones 1991).
Accurate solutions to the CR diffusion-convection equation
require a computational grid spacing significantly smaller than $D_{\rm diff}$,
typically, $\Delta x \sim 0.1 D_{\rm diff}(p)$.
On the other hand, for a realistic diffusion transport model with a steeply
momentum-dependent diffusion coefficient,
the highest energy, relativistic particles have diffusion lengths
many orders of magnitude greater than those of the lowest energy particles.

To follow the acceleration of highly relativistic CRs from suprathermal
energies, all those scales need to be resolved numerically.
However, the diffusion and acceleration
of the low energy particles are important only close to the shock owing
to their small diffusion lengths. 
Thus it is necessary to resolve numerically the diffusion length of the
particles only around the shock.
To solve this problem generally we have developed the CRASH code 
by combining a powerful ``Adaptive Mesh Refinement'' (AMR) technique
(Berger \& Le Veque 1998) and a ``shock tracking''
technique (Le Veque \& Shyue 1995), and implemented them into a hydro/CR code
based on the wave-propagation method (Kang \etal 2001; Kang \etal 2002).
The AMR technique allows us to ``zoom in'' inside the precursor structure
with a hierarchy of small, refined grid levels applied around the shock.
The shock tracking technique follows hydrodynamical shocks within
regular zones and maintains them
as true discontinuities, thus allowing us to refine the region around
the gas subshock at an arbitrarily fine level.
The result is an enormous savings in both computational time and
data storage over what would be required to solve the problem using
more traditional methods on a single fine grid.

\subsection{Injection Model}
In the ``thermal leakage'' injection model, 
some suprathermal particles in the tail of the Maxwellian distribution 
swim successfully against the Alfv\'en waves advecting downstream, and then 
leak upstream across the subshock and get injected in the CR population.
In order to model this injection process 
in Gieseler \etal (2001) 
we adopted a ``transparency function'',
$\tau_{\rm esc}$, which expresses the probability that supra-thermal
particles at a given velocity can leak upstream through the magnetic waves,
based on non-linear particle interactions with self- generated waves
(Malkov and V\"olk 1998).
In this scheme, the transparency function is {\it approximated} by
the following functional form,
\begin{eqnarray}
\tau_{\rm esc}(\epsilon,\upsilon/u_d)~=~H[ \tilde{\upsilon}-(1+\epsilon)]
        (1-\frac{u_d}{\upsilon})^{-1}\,
        (1-\frac{1}{\tilde{\upsilon}})\nonumber\\
 \cdot\, \exp\{-[\tilde{\upsilon}-(1+\epsilon)]^{-2}\},
\label{transp}
\end{eqnarray}
which depends on the ratio of particle velocity, $\upsilon$,
to downstream flow velocity in the subshock rest-frame,
$u_d$.  Here $H$ is the Heaviside step function.
The inverse wave-amplitude parameter, $\epsilon = B_0/B_{\perp}$, is defined
in Malkov and V\"olk (1998) and measures the ratio of the amplitude of the
postshock MHD wave turbulence $B_{\perp}$ to the general magnetic field 
aligned with the shock normal, $B_0$.
Here $\tilde{\upsilon}= \epsilon ~\upsilon/u_d$ is the normalized
particle velocity.
This function behaves like a smoothed step function, and the injection
takes place in momentum space where the $\partial \tau_{\rm esc} / \partial p$ 
is greatest.
The breadth of the thermal velocity distribution relative the
downstream flow velocity in the subshock rest-frame (\ie $\upsilon_{\rm th}/u_d$)
determines the probability of leakage, and so the injection process is sensitive 
to the velocity jump at the subshock, which depends 
on the subshock Mach number.
The injection rate increases with the subshock Mach number, but becomes
independent of $M_s$ in the strong shock limit of
$M_s\gsim 10$ (Kang \etal 2002).
Thus the injection rate should change with the shock strength in the 
time-dependent evolution of CR modified shocks, 
as the subshock weakens by way of precursor compression.

The only free parameter of the adopted transparency function is
$\epsilon$ and it is rather well constrained, since $0.3\lsim \epsilon \lsim 0.4$ 
is indicated for strong shocks (Malkov \& V\"olk 1998).
However, it turns out the injection rate depends sensitively on the value of $\epsilon$,
due to the exponential cut off in a thermal velocity distribution.
It is also expected that the wave generation is weaker for low Mach shocks,
leading to the larger values of $\epsilon$.
So in this study we will consider $0.2 \le \epsilon \le 0.4$. 

In the CRASH code, in order to emulate numerically thermal leakage injection,
we first estimate the number of suprathermal particles
that cross the shock according to the diffusion-convection equation,
and then we allow only a small fraction of the combined advective and diffusive
fluxes to leak upstream with the probability prescribed by $\tau_{\rm esc}$.
The readers are referred to Kang \etal (2002) for more details of our 
numerical scheme for thermal leakage injection model.
\begin{figure*}[t]
\vskip -0.5cm
\centerline{\epsfysize=14cm\epsfbox{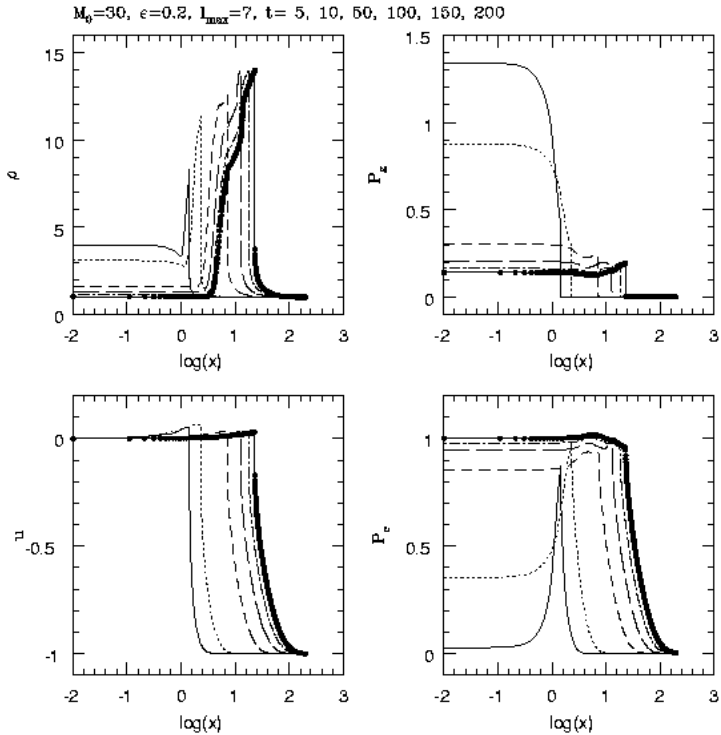}}
\vskip -0.5cm
\caption{
Time evolution of the shock driven by an accretion flow with $\rho_0=1$,
$u_0=-1$ and $M_0=30$ which is reflected at $x=0$.
Seven levels of refinements ($l_{\rm max}=7$) were used and the
inverse wave-amplitude parameter $\epsilon=0.2$ was adopted.
The snapshots are shown at $t =$ 5, 10, 50, 150, and 200.
The shock propagates to the right, so the leftmost profile corresponds
to the earliest time.
For $t =$ 200, data at each cell is shown as filled circles
to show clearly the subshock jump.
Note the distance from the reflecting plane is in a logarithmic scale. 
}
\end{figure*}

\begin{figure*}[t]
\vskip -0.5cm
\centerline{\epsfysize=14cm\epsfbox{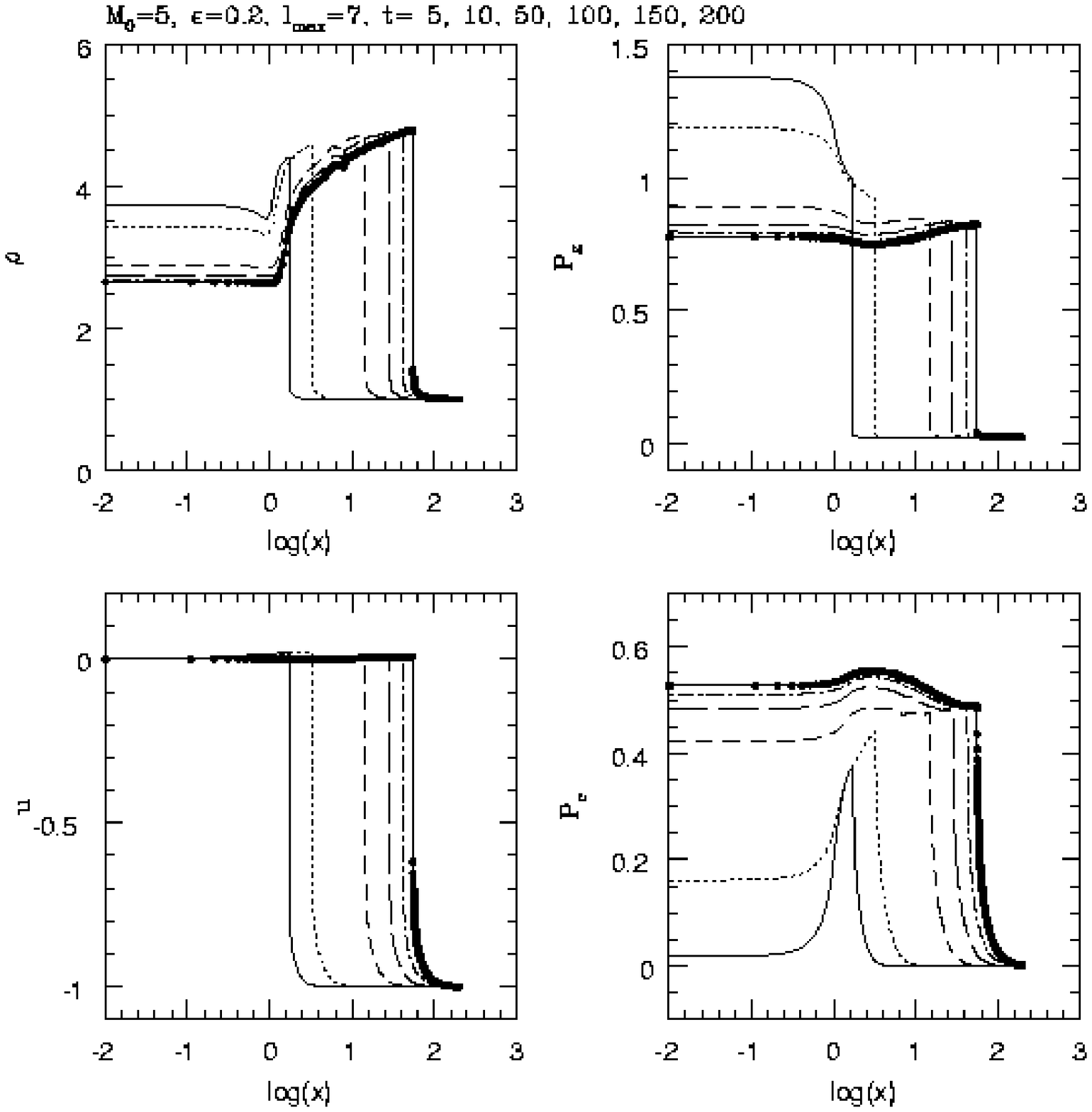}}
\vskip -0.5 cm
\caption{
Same as Figure 1 except $M_0=5$.
}
\end{figure*}
\section{Numerical Model}
\subsection{Diffusion Coefficient Model}

Diffusive acceleration at shocks depends on the
existence of Alfv\'enic turbulence capable of strongly scattering 
energetic protons.
The irregularities in the magnetic field within the cluster 
can be generated by various dynamical effects 
such as the supersonic motion of galaxies through ICM, 
mergers of substructures and galactic winds.
In addition, according to the cosmological hydrodynamic simulations,
the downstream regions of the accretion shocks (\ie ICM) are full of
turbulent flow motions. 
Some observations (Feretti \etal 1995) indicate that the ICM field is
likely to be tangled on scales of the order of less than 1 kpc.
Then turbulent flows may subsequently generate the irregularities in
magnetic field.
Outside the accretion shock, however, we must hypothesize the
existence of pre-existing field turbulence.
Lucek \& Bell (2000) have shown by numerical simulations that 
CR streaming induces large-amplitude Alfv\'en waves in a quasi-parallel
shock, implying that the shock formation
process itself can generate the necessary turbulent field.
So it is commonly assumed in calculations of diffusive shock
acceleration that the scattering Alfv\'en waves are self-generated by the
CRs themselves by the so-called ``streaming instability'' (Wentzel 1974).
Adiabatic compression of that field and turbulence by convergence
of the inflow onto the accretion shock might very reasonably lead
to an acceptable level of magnetic irregularities in the shock
vicinity.

The Bohm diffusion model represents a saturated wave spectrum
and gives the minimum diffusion coefficient
as $\kappa_{\rm B} = 1/3 r_{\rm g} \upsilon$,
when the particles scatter within one gyration radius ($r_{\rm g}$)
due to random scatterings off the self-generated waves.
This gives
\begin{equation}
\kappa(p) = \kappa_{\rm o} {{p^2}\over {(p^2+1)^{1/2}}},
\end{equation}
where
$\kappa_{\rm o}= 3.13\times 10^{22} {\rm cm^2s^{-1}} B_{\mu}^{-1}$
and $B_{\mu}$ is the magnetic field strength
in units of microgauss.
In order to model amplification of self-generated turbulent waves
due to compression of the perpendicular component of the magnetic field,
the spatial dependence of the diffusion is modeled as
\begin{equation}
\kappa(x,p) = \kappa(p)({\rho_0 \over \rho(x)}),
\end{equation}
where $\rho_0$ is the upstream gas density.
This form is also required to prevent the acoustic instability
of the precursor (Kang, Jones \& Ryu 1992).
 
\subsection{Time and Length Scales}

The mean acceleration time scale for a particle to reach momentum
$p$ is determined by the velocity jump at the shock and the
diffusion coefficient (e.g., Drury 1983), that is,
\begin{equation}
\tau_{acc}= {p\over<{dp \over dt}>} 
= {3\over {u_1-u_2}} ({\kappa_1\over u_1} + {\kappa_2\over u_2}).
\end{equation}
Here the subscripts, 1 and 2, designate the upstream and downstream
conditions, respectively.
If the strong shock limit is taken (\ie $u_2=u_1/4$), and we assume for a
turbulent field that $B/\rho$ is constant across the shock (\ie
$\kappa/u = {\rm constant}$), then
$\tau_{acc} \approx 8 \kappa_{\rm B} / u_s^2 $. 
Using the Bohm diffusion coefficient in equation (9), the mean
acceleration time scale is given by
\begin{equation}
\tau_{acc} \approx (7.92\times 10^9 {\rm years})
({p \over 10^{10}}) B_{\mu}^{-1} ({u_s \over 1000 {\rm km~s^{-1}}})^{-2}.
\end{equation}
This shows that the diffusive acceleration time scale increases directly
with the particle energy and that it takes about a billion years to accelerate
the proton to $E=10^{18} $eV at a typical cluster shock,
assuming $B_{\mu} \sim 1$.
The diffusion length of the CR protons is given by
\begin{equation}
D_{\rm diff}= {\kappa_{\rm B} \over u_s}\nonumber\\
= 1.02 {\rm Mpc} ({p \over 10^{10}}) B_{\mu}^{-1} ({u_s \over 1000 {\rm km~s^{-1}}})^{-1}.
\end{equation}
So the CR protons of $E=10^{18}$eV  diffuse on the length scale 
comparable to the cluster size.
Note the mildly relativistic protons ($p\sim 1-10$) are almost instantaneously
accelerated at these shocks compared to the cosmological time scale
and diffuse on the length scale much smaller than the cosmological scale. 

\begin{figure*}[t]
\vskip -0.5cm
\centerline{\epsfysize=14cm\epsfbox{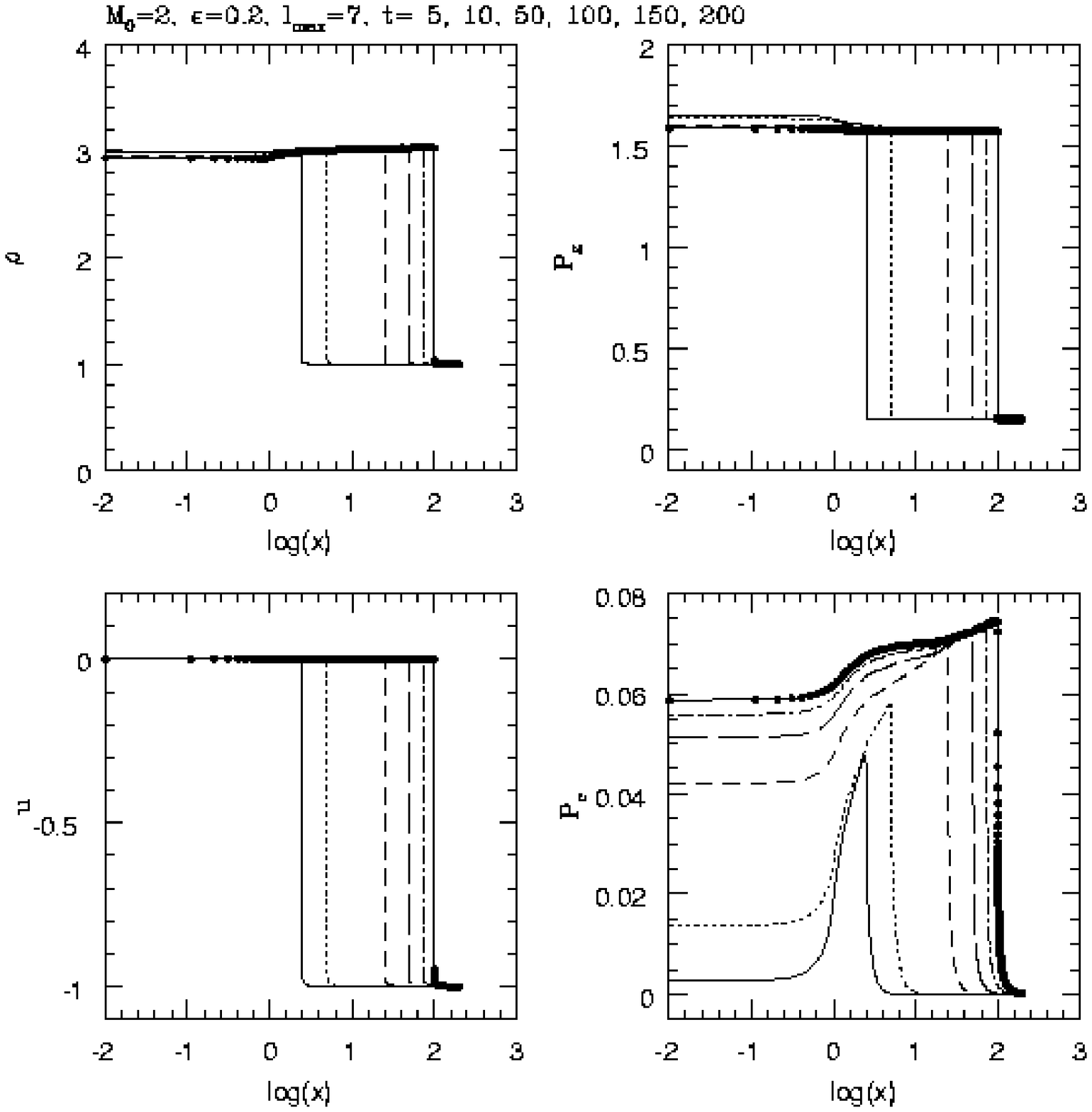}}
\vskip -0.5 cm
\caption{
Same as Figure 1 except $M_0=2$.
}
\end{figure*}
\begin{figure*}[t]
\vskip -0.5cm
\centerline{\epsfysize=14cm\epsfbox{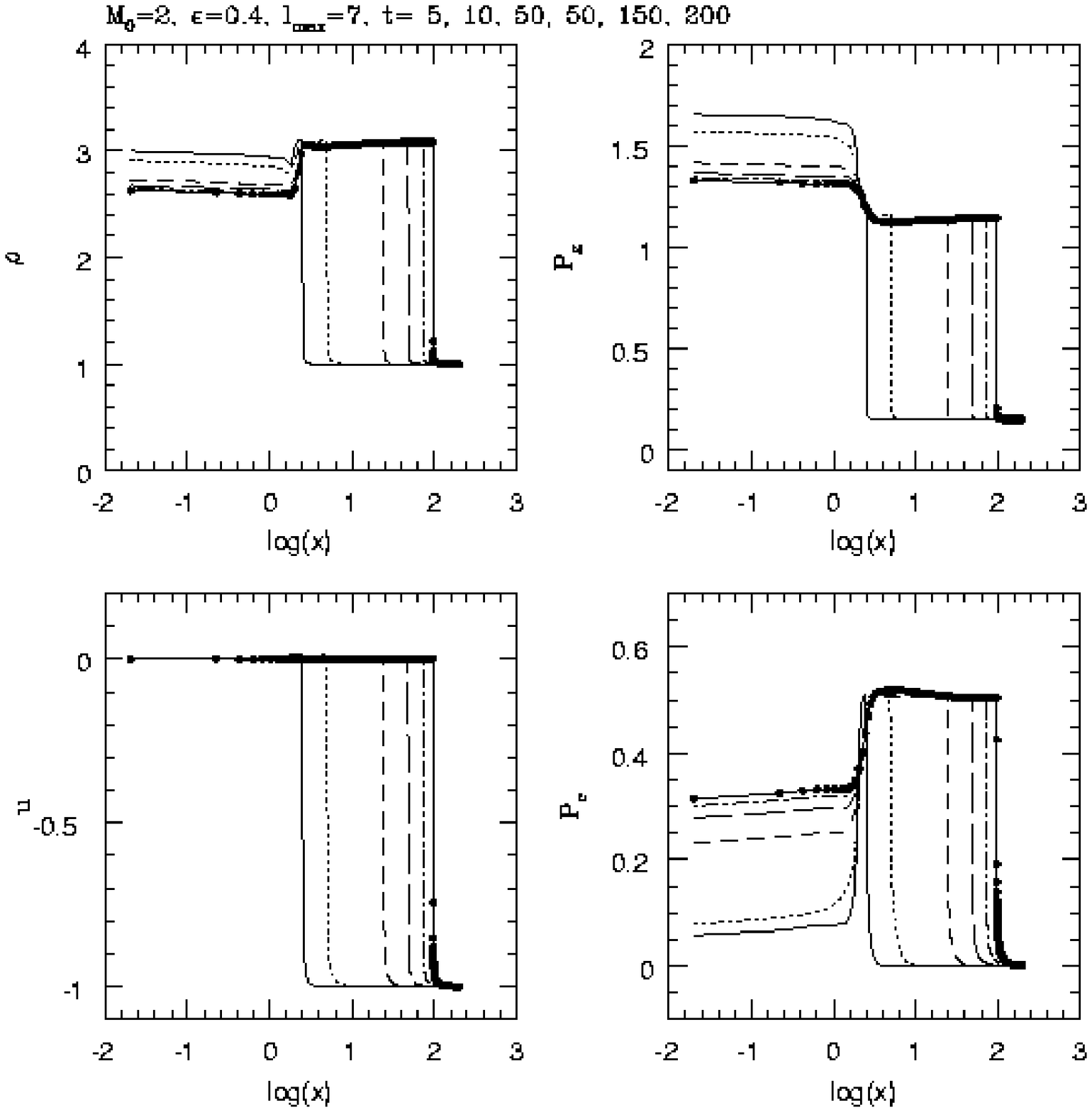}}
\vskip -0.5 cm
\caption{
Same as Figure 1 except $M_0=2$ and $\epsilon=0.4$.
}
\end{figure*}
\subsection{1D Plane-parallel Shock Models}

In this contribution, we study the CR acceleration and its dynamical
effects at one-dimensional (1D) quasi-parallel shocks which form due to 
the accretion flows in a plane-parallel geometry. 
In general, cosmic shocks associated with large scale structure formation
can be oblique and have various geometries, depending on the types of nonlinear
structure onto which the accretion flow falls. Roughly speaking we have: 
1D plane-parallel shocks around the sheets, 
2D cylindrical shocks around the filaments, 
and 3D spherical shocks around the clusters of galaxies. 
Due to severe requirements on the computational resources, our simulations
can follow the acceleration of the protons from suprathermal energies 
($p\sim 10^{-3}$) to mildly relativistic energies ($p\sim 50$).
For the particles in this energy range, the acceleration time scales ($< 40$ years) 
are much shorter than and the cosmological time scale ($t_{\rm H}\sim 10^{10}$ years)
and the diffusion length scales ($<0.1$pc) are much smaller than the
curvature of multi-dimensional cosmic shocks ($R_s\sim $ a few Mpc).
On those scales 
where $D_{\rm diff} \ll R_s$, 
the diffusion and acceleration of CRs can be studied with
the 1D plane-parallel shock models. 
Thus, for our purposes the shocks that occur in major mergers of substructures can be represented  
approximately by 1D plane-parallel shocks.
As shown in Kang \etal (2002), the maximum CR pressure 
and the shock structure
modification approach to time asymptotic limits as the particles become 
relativistic $p \sim 1$.
This is due to the balance between acceleration and diffusion of CRs,
that is,  fewer particles are accelerated to higher energies,
and higher energy CRs diffuse over larger volume of space as the CR shock evolves.
Thus our simulations can predict long-term behaviors of CR modified
shocks, even though our integration time is only a small fraction of $t_H$.

Here we ignore gravity, so changes in the infall velocity
and density, and the background cosmological evolution, because
the integration time of our simulations is much shorter than any of the cosmological time scales. 
So we consider a pancake shock formed by the steady accretion flow with
a constant density and pressure:  
a 1D simulation box with [0,$x_{max}$] and
an accretion flow entering into the right boundary of the
simulation box with a constant density,
$\rho_0$, pressure, $P_{\rm g,0}$, and velocity, $u_0$.
The flow is reflected at the left boundary ($x=0$) (\ie pancake middle plane). 
A shock forms and propagates to the right. 
For a hydrodynamic shock without the CRs, the shock speed
is $u_s = |u_0|/(r-1)$ in the simulation frame
and $u_s^\prime=|u_0|r/(r-1)$ in the far upstream rest frame,
where $r$ is the compression ratio across the shock.
Throughout the paper we denote the velocities in the simulation frame
by the unprimed variables, while the velocities in the rest frame of far 
upstream flow by the primed variables.
For strong hydrodynamics shocks, $r=4$ and $u_s^\prime=(4/3)|u_0|$.
A CR modified shock consists of a smooth precursor and a subshock,
since the CRs diffuse upstream of the subshock, and the CR pressure
decelerates and heats the preshock flow adiabatically,
resulting in weakening of the subshock.
We denote the values immediately upstream of the
subshock as $u_1$, $\rho_1$, and $P_{\rm g,1}$ and 
the values immediately downstream of the subshock as $u_2$, $\rho_2$ and
$P_{\rm g,2}$.
We also denote the subshock speed relative to the immediate upstream
flow as $u_{\rm sub}$, which is smaller than $u_s^\prime$ due to the precursor
deceleration. 

We designate the shock models by
the Mach number of the accretion flow that is defined by 
\begin{equation}
M_0 = { |u_0| \over (\gamma P_{\rm g,0}/\rho_0)^{1/2}},
\end{equation}
where once again $u_0$, $\rho_0$, $P_{\rm g,0}$ are the accretion
velocity, the density and pressure of the infall flow, respectively.
The Mach number of the accretion shock resulting from such accretion flow
is $M_s \approx 4/3 M_0$ for strong gasdynamic shocks. 
We set the far upstream density and flow values as $\rho_0=1$, $u_0=-1$ 
in code units for all models, while we vary 
the gas pressure, $P_{\rm g,0}$ for different values of $M_0$,
where $ 2 \le M_0 \le 100$.
So the preshock gas is colder for high values of $M_0$.

The following parameters are used:
the gas adiabatic index, $\gamma_{\rm g}=5/3$ and the velocity normalization
constant $u_{\rm o}=1500~{\rm km s}^{-1}$ ($\beta=u_{\rm o}/c = 0.005$).
Then the diffusion coefficient, $\kappa_{\rm o} =3.13\times 10^{21}
{\rm cm^2s^{-1}}B_{\mu}^{-1}$, determines the length and time scales
as $x_{\rm o} = \kappa_{\rm o}/u_{\rm o}$ and
$t_{\rm o}=\kappa_{\rm o}/u_{\rm o}^2$, respectively, which represent,
in fact, the diffusion length and time scales for the protons of $p=1$. 
For $B_{\mu}=1$, 
$x_{\rm o} = 2.1\times 10^{14}$ cm and $t_{\rm o}= 1.4 \times 10^6 $s.
These scales are much smaller than cosmological length and time scales as
discussed before, which justifies our assumptions about the steady accretion
flow and the 1D plane-parallel geometry.
The gas density and pressure normalization constants, $\rho_{\rm o}$ and 
$P_{\rm o}=\rho_{\rm o} u_{\rm o}^2$, are arbitrary.
For the hydrogen number density of $n_H=10^{-4}{\rm cm^{-3}}$, for example, 
$\rho_{\rm o} = 2.34 \times 10^{-28} {\rm g~cm^{-3}}$ and
$P_{\rm o}= 5.26 \times 10^{-12} {\rm erg~cm^{-3}}$.
So if one assumes the Bohm diffusion model, the three parameters,
\ie $\beta$, $B_{\mu}$, and $M_0$, sufficiently define a CR shock model.
Throughout the paper and in the code  
physical variables are given in units of the normalization
constants, $x_{\rm o}$, $t_{\rm o}$, $u_{\rm o}$, $\rho_{\rm o}$,
and $P_{\rm o}$.

Although the theoretically preferred values of the inverse wave-amplitude 
parameter, $\epsilon$, lie between
0.3 and 0.4 for strong shocks (Malkov 1998),
such values lead to very efficient initial injection and most of the shock
energy is transfered to the CR component for strong shocks of 
$M_0 \gsim 30$ (Kang \etal 2002).
As a more conservative option we have considered a set of models for
$2\le M_0 \le 100$ with $\epsilon=0.2$ and
another set of models for $M_0=2$ with $0.2 \le \epsilon \le 0.4$.  
The former is chosen to explore the dependence of the CR acceleration
on the accretion flow Mach number for a given value of $\epsilon$, while
the latter is chosen to explore the dependence on the injection rate for
a low Mach number flow.

The simulations were carried out on a base grid with
$\Delta x_0 = 0.02$ using $l_{\rm max}=7$ additional grid levels,
so $\Delta x_7 = 1.56\times 10^{-4}$ on the finest grid.
The simulated space is $x=[0,200]$ and $N=10000$ zones are used
on the base grid for the models that were integrated to $t/t_o=200$.
For the models that were integrated to $t/t_o \lsim 100$, $x=[0,100]$
and $N=5000$ are used.
The number of refined zones around the shock is $N_{rf}=100$ on the base
grid and so there are $2N_{rf}=200$ zones on each refined level.
The length of the refined region at the base grid is 2, so $1/100$ of the
entire simulated space on the base grid is refined.
To avoid technical difficulties,
the multi-level grids are used only after the shock propagates away from
the left boundary at the distance of $ x_s = 1$.
With $x_s <1$, the downstream refined region is outside the left boundary
of the simulation box and the full length of the refined region around 
the shock cannot be set down with the current version of the CRASH code.
After the shock moves to $x_s=1$ (at $t \approx 3 $ for strong shocks),
the AMR technique is used and 
the CR injection and acceleration are activated.
This initial delay of the CR injection and acceleration should not
affect the final outcomes. 
We integrate the simulation until $ t=100-200$,
so that the maximum momentum achieved by the end of simulation is of order 
of $p_{\rm max} \sim 40$, above which the CR distribution function decreases 
exponentially.
For all models we use 230 uniformly spaced logarithmic momentum zones
in the interval $\log (p/m_p c)=[\log p_0,\log p_1]=[-3.0,+3.0]$

For gasdynamic variables the continuous boundary condition 
is used at right boundary,
while the reflecting boundary condition is applied to left boundary 
of the simulation box.
For the CR distribution function a continuous boundary is assumed for
the advection step and a no-flux boundary condition is adopted
for the spatial diffusion step.
Either below $p_0$ or above $p_1$; $g(p)=0$ is assumed.

\section{RESULTS}

\subsection{Modified Shock Structure}
We show the time evolution of shocks at $t=5,$ 10, 50, 100, 150 and 200
for models with the accretion flow Mach number, $M_0= 30,$ 5, and 2 in Figures 1-3. 
The inverse wave amplitude parameter was assumed to be $\epsilon=0.2$. 
As CRs are injected and accelerated at the shock, 
the CR pressure increases and diffuses upstream, leading to
a precursor in which the upstream flow is decelerated and compressed
adiabatically.
As the CR precursor grows, the subshock slows down and the postshock
density increases, while the postshock gas pressure decreases.
Because the injection rate is quite high for strong shocks,
the CR energy increases and the modification
to the flow structure proceeds very quickly in a time scale
comparable to $t_{\rm acc}$ for $p\sim 2-3 $. 

Kang \etal (2002) also considered similar shock models, but
with different flow conditions at the left boundary. 
There a gasdynamic shock was set up initially and allowed to evolve 
with open boundaries both far upstream and downstream.
Because the CRs are not allowed to diffuse out at the reflecting,
downstream boundary in the current models, 
the CR pressure builds up faster.   
Otherwise, however, the CR shocks evolve similarly in both
models:
1) The total transition consists of a precursor and a subshock
that weakens to a lower Mach number shock, but does not disappear entirely. 
2) After an initial quick adjustment, the CR pressure at the shock reaches
{\it approximate} time-asymptotic values when the fresh injection
and acceleration are balanced with advection and spreading of high 
energy particles due to strong diffusion.
For strong shocks,  $P_{\rm c,2}/\rho_0 (u_{s,0}^{\prime}) ^2 \rightarrow 0.56$,
where $u_{s,0}^{\prime}$ is the shock speed before any significant nonlinear
CR feedback occurs.
3) Once the postshock CR pressure becomes constant,
the shock structure evolves approximately in a ``self-similar'' way,
because the scale length of shock broadening increases linearly with 
time.
4) A postshock ``density spike'' forms due to the nonlinear feedback
of the CR pressure  (see also Jones \& Kang 1990). 
5) For a given inverse wave-amplitude parameter, $\epsilon$, the CR acceleration
efficiency and the flow modification depend sensitively on the
shock Mach number, but they seem to converge at the strong shock 
limit ($M_s \gsim30$).

For the $M_0=30$ model, for example,  the initial unmodified shock speed
is $u_{s,0}^\prime = 4/3$ (Mach number $M_s=40$), but the shock
slows down to $u_{s}^\prime \approx 1.1$ due to CR nonlinear modification.
So the ratio $P_{\rm c,2}/\rho_0 (u_s^{\prime}) ^2 \rightarrow 0.8$,
where $u_{s}^\prime$ is the instantaneous shock speed.
For strong shocks 
the CR pressure dominates over the gas pressure with a 
CR injection fraction $\sim 10^{-4}-10^{-3}$ (see Figure 8 below),
and a ratio $P_{\rm c,2}/\rho_0 (u_s^{\prime}) ^2 \rightarrow 0.8-1.0$,
which is consistent with the simulation results of
Kang \etal (2002).
For the $M_0=5$ model, $P_{\rm c,2}/\rho_0 (u_{s,0}^{\prime}) ^2 \rightarrow 0.27$
and still shows substantial degree of the flow modification.  
For $M_0=2$ model, however, the CR pressures increases only to
$P_{\rm c,2}/\rho_0 (u_s^{\prime}) ^2 \sim 0.03$ and the flow
structure is not modified at all.
According to Miniati \etal (2000), the cosmic shocks inside the ICM
(\ie merger shocks) have mostly low Mach numbers of $ 1 \lsim M_s \lsim 5$, 
because these shocks propagate into the already hot medium. 
But the shock speed is not much different from the accretion
shocks that propagate into the cold medium and so have high Mach numbers.
Thus the sample models shown in Figures 1-3 can provide qualitative pictures
on the effects of the CR acceleration to dynamics of cosmological shocks. 

We note, however, that the value of $\epsilon=0.2$ is perhaps  
unrealistically small for a low Mach number shock, 
since the self-generation of waves is expected to be less efficient. 
So, we considered three additional models for $M_0=2$ ($M_s=3$) with larger 
values of $\epsilon=0.25$, 0.3, and 0.4. 
The CR pressure increases with higher values of $\epsilon$ due to higher 
leakage fractions, as expected.  So, for example, 
the ratio $P_{\rm c,2}/\rho_0 (u_s^{\prime}) ^2 \sim 0.22$ for $\epsilon=0.4$
model (see Figure 4).
But even in this rather high injection model with a CR fraction of $\xi \sim
10^{-2}$, the flow modification is minimal except for the reduction of the gas pressure.
As expected, the CR acceleration is very inefficient 
and the CR nonlinear feedback is insignificant   
in weak shocks of a Mach number of a few. 

\begin{figure*}[t]
\vskip -0.5cm
\centerline{\epsfysize=14cm\epsfbox{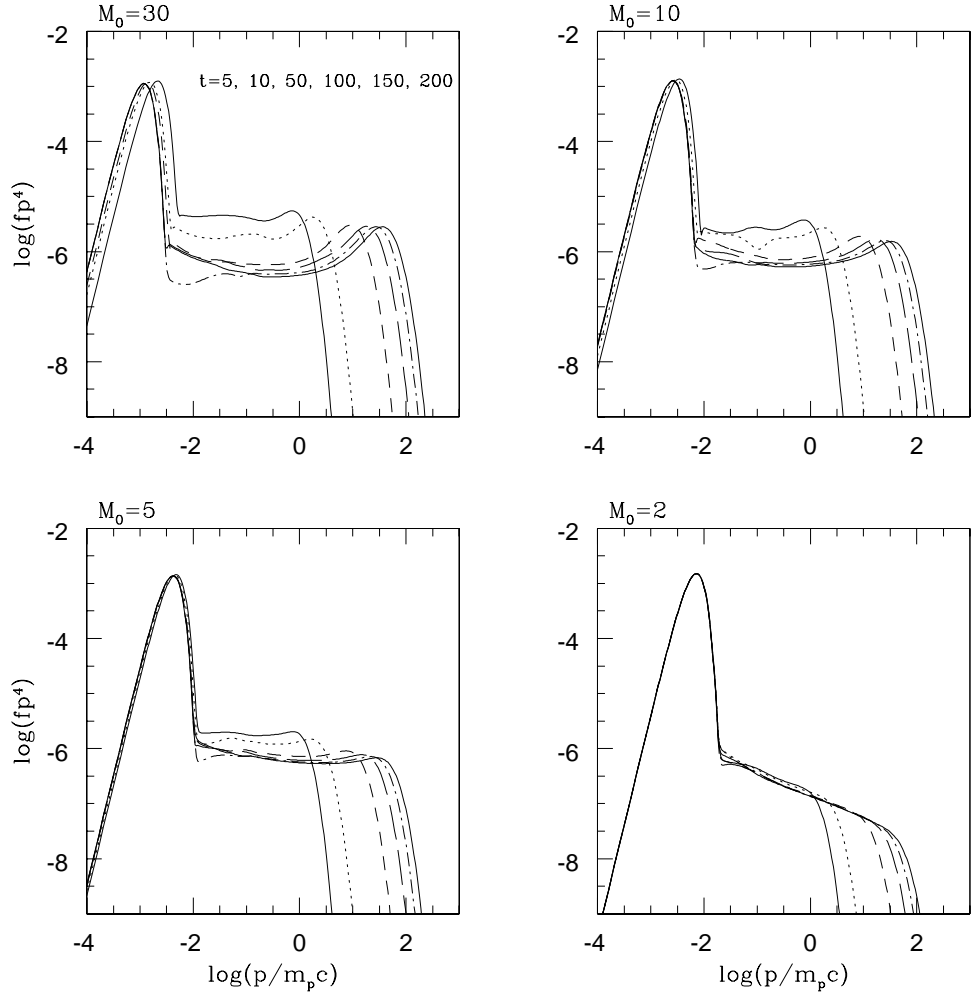}}
\vskip -0.5cm
\caption{
Evolution of the CR distribution function at the shock, represented
as $g=p^4f(p)$, is plotted for 
the models of $M_0=$ 30, 10, 5, and 2 with $\epsilon=0.2$.
}
\end{figure*}
\begin{figure*}[t]
\vskip -0.5cm
\centerline{\epsfysize=14cm\epsfbox{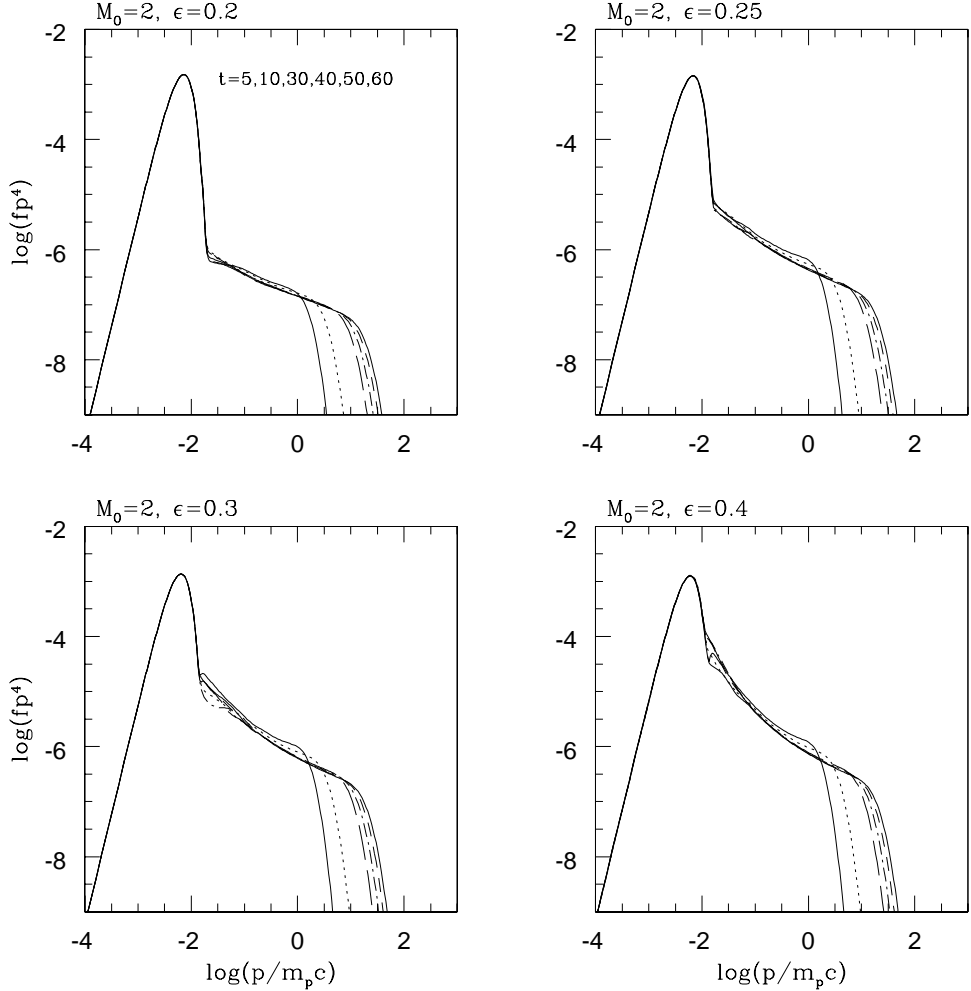}}
\vskip -0.5cm
\caption{
Evolution of the CR distribution function at the shock, represented
as $g=p^4f(p)$, is plotted for the $M_0=2$ models with 
$\epsilon=0.2,$ 0.25, 0.3 and 0.4.
}
\end{figure*}

\subsection{Particle Distribution Function}

\begin{figure*}[t]
\vskip -0.5cm
\centerline{\epsfysize=15cm\epsfbox{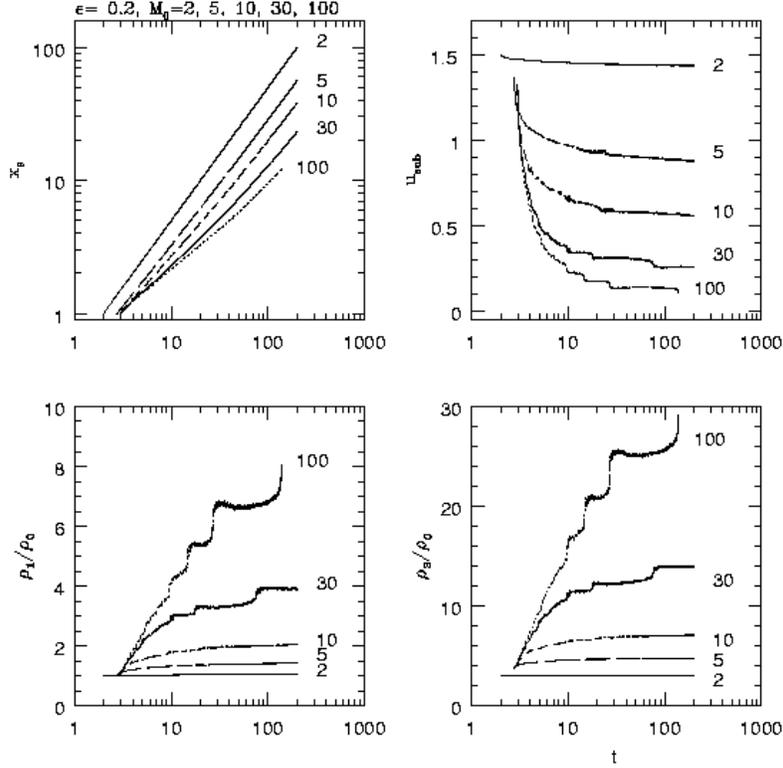}}
\vskip -0.5cm
\caption{
Shock location from the reflecting plane, $x_s$,
the shock velocity relative to the immediate preshock flow, $u_{\rm sub}$,
preshock density, $\rho_1/\rho_0$, and
postshock density, $\rho_2/\rho_0$ are plotted as a function of time
for $M_0=2-100$ and $\epsilon=0.2$.
}
\end{figure*}

\begin{figure*}[t]
\vskip -0.5cm
\centerline{\epsfysize=15cm\epsfbox{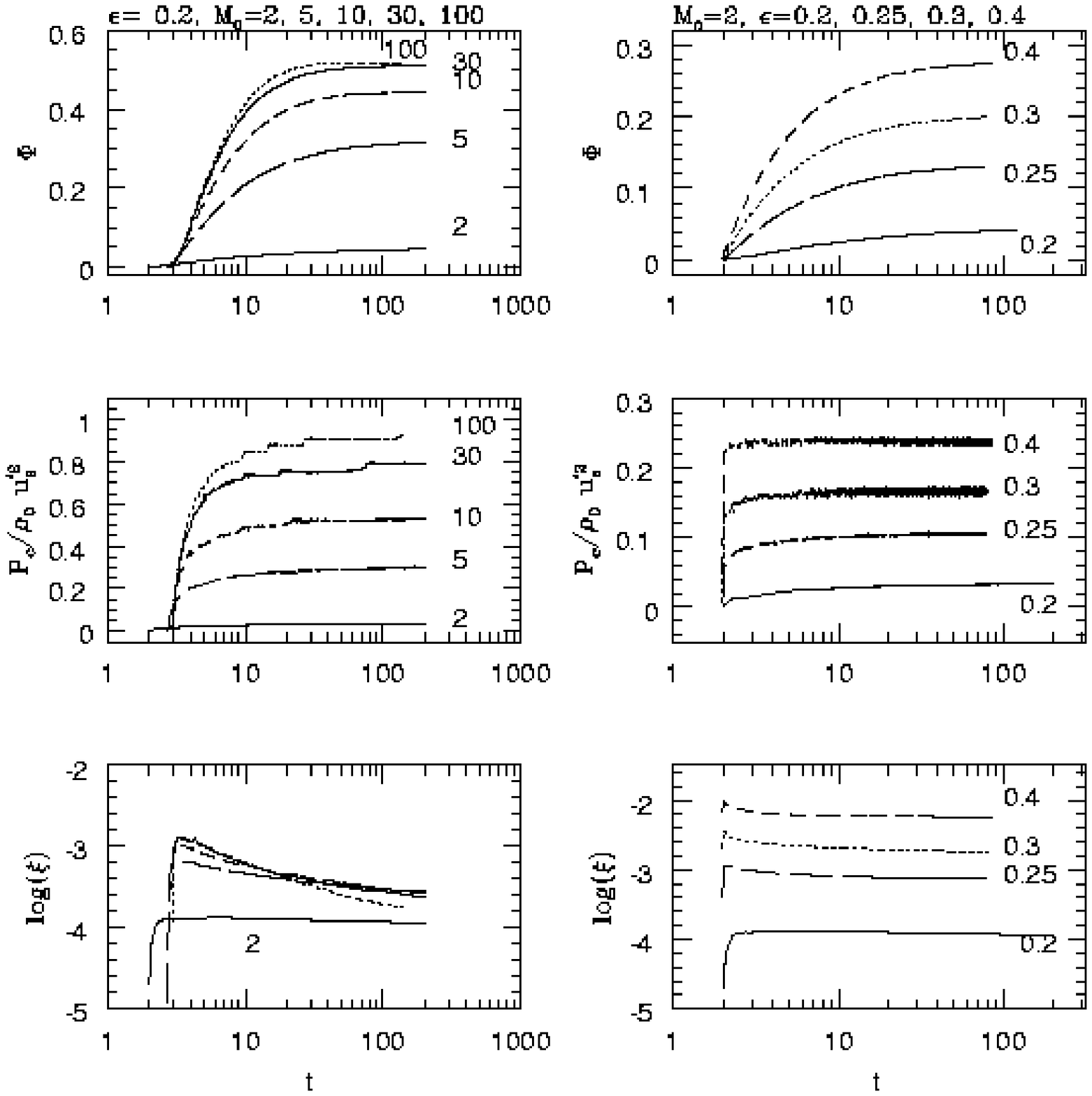}}
\vskip -0.5cm
\caption{
The ratio of total CR energy in the simulation box
to the kinetic energy in the initial shock rest frame
that has entered the simulation box from upstream, $\Phi(t)$,
the postshock CR pressure in units of far upstream ram pressure in the 
instantaneous shock frame, 
and time-averaged injection efficiency, $\xi(t)$.
Left three panels are for $M_0=2-100$ and  $\epsilon=0.2$.
Right three panels show the same quantities for $M_0=2$ and  
$\epsilon=0.2$, 0.25, 0.3, and 0.4.
}
\end{figure*}

Evolution of the CR distribution function at the shock 
is given for the models of $M_0 = 30, 10, 5,$ and 2
with $\epsilon=0.2$ in Figure 5.  
Thermal particles are represented by a Maxwellian distribution below
$ p\sim 10^{-2}$.
One can see that the Maxwellian distribution shifts to lower momenta,
as the postshock temperature reduces due to energy transfer to CRs. 
Just above the injection pool, the distribution function changes smoothly
from the thermal distribution to an approximate power-law whose index is
close to the test--particle slope for the subshock.
While a fraction of particles injected earlier continues to be
accelerated to higher momenta, so that $p_{\rm max}(t)$ increases,
the amplitude of $g(p)$ at the shock and at a given momentum
decreases with time for $p<p_{\rm max}(t)$ due to diffusion.
The distribution function $g(p)$ shows the
characteristic ``concave upwards'' curves reflecting 
modified shock structure (including the precursor)
for the shocks with $M_0\ge 10$.
For the models with $M_0 = 2$ and $\epsilon=0.2$, 
the CR modification to the flow structure is insignificant,
so the particle distribution is a power-law with the test-particle
slope (\ie $f(p) = f_0 p^{-q}$ and  $q = 3r/(r-1)\approx 4.5$, 
where $r=3$ for $M_s=3$). 

In Figure 6, the CR distribution function at the shock is shown
for the $M_0=2$ model for $\epsilon=0.2$, 0.25, 0.3, and 0.4.
For weaker wave fields (larger $\epsilon$) the particles with lower
momenta can leak upstream, so the injection pool is located at 
the lower momenta closer to the peak of the Maxwellian distribution,
resulting in a higher injection rate.
Even though about 20 \% of the shock kinetic energy is transfered
to the CR component for $\epsilon=0.4$, the flow velocity structure is 
only slightly affected.
So the CR distribution function is dictated by the test-particle
like power-law spectrum for $M_0=2$ models, roughly independent of the
value of $\epsilon$. 

\subsection{Subshock Evolution}
The speed of the ``initial hydrodynamic shock'', $u_{s,0}^{\prime}$, 
that emerges from the reflecting plane before the CR injection/acceleration 
begins depends on the Mach number of the accretion flow. 
It is given by $u_{s,0}^{\prime} = r/(r-1) |u_0|$, so that
$u_{s,0}^{\prime}= 1.5$ for $M_0=2$,
$u_{s,0}^{\prime}= 1.36$ for $M_0=5$, 
and $u_{s,0}^{\prime}= 4/3 $ for $M_0\ge 10$.
After the CR injection/acceleration starts, the CR pressure modifies
the shock flow and the ``instantaneous shock'' speed relative to the 
far upstream flow, $u_s^{\prime}$, decreases.
We plotted the position of the shock, $x_s$, against time
for models with $\epsilon = 0.2$ in the upper left panel of Figure 7.  
We see that the shock is decelerated further in higher Mach number models,
since the CR pressure is greater and so is its dynamical feedback to the flow. 

The injection rate in the thermal leakage injection model depends on the
strength of the subshock, so the subshock speed relative to the immediate
preshock gas in the precursor, $u_{\rm sub}$, is important.
Because of the pre-deceleration in the precursor, 
especially for strong shocks, $u_{\rm sub}$ can be
much smaller than the shock speed relative to the far upstream.
We plotted this subshock speed in the upper right panel of Figure 7.
For $M_0=2$ $u_{\rm sub} \approx u_s^{\prime}$, but
$u_{\rm sub} \ll u_s^{\prime}$ for $M_0 \ge 30$.
We also plotted the pre-subshock gas density, $\rho_1/\rho_0$, 
and the immediate postshock gas density, $\rho_2/\rho_0$,
in the lower two panels. 
The modification to the flow velocity occurs mostly before $ t \lsim 10$, 
so the subshock velocity decreases at the same time, but
the precursor compression continues even after $t>10$, 
especially in the strong shocks. 
We note that the subshock persists and so a completely smooth 
transition never develops by the termination time of our simulations.

\subsection{Injection and Acceleration Efficiencies}

We define the injection efficiency as the fraction of particles that have
entered the shock from far upstream and then injected into the CR distribution: 
\begin{equation}
\xi(t)=\frac {\int_0^{x_{max}} {\rm d x} \int_{p_0}^{p_1} 4\pi f_{\rm CR}(p,x,t)p^2 {\rm d p}}
{ \int_{t_1}^t n_0 u_s^\prime (t^{\prime}) {\rm d t^{\prime}} }\, 
\end{equation}
where $f_{\rm CR}$ is the CR distribution function, 
$n_0$ is the particle number density far upstream,
and $t_1$ is the time when the CR injection/acceleration is turned on
($t_1 \approx 3$ for $M_0\ge 5$ and $t \approx 2$ for $M_0=2$).
If the subshock becomes steady, then $n_0 u_s^\prime$ is the same as the
particle flux swept by the subshock, $n_1 u_{\rm sub}$, where 
$n_1$ is the particle number density immediately upstream to the subshock.
In our simulations these two fluxes can differ up to 10 \% for strong
shock models, because the shock speed is changing slowly.

As a measure of acceleration efficiency,
we define the  ``CR energy ratio''; namely the
ratio of the total CR energy within the simulation box to
the kinetic energy in the {\it initial shock frame} 
that has entered the simulation box from far upstream,
\begin{equation}
\Phi(t)=\frac {\int_0^{x_{max}} {\rm d x} E_{\rm CR}(x,t)}
 {  0.5\rho_0 (u_{s,0}^{\prime})^3  t }.
\label{crenrat}
\end{equation}
Since our shock models have the same accretion density and velocity, but
different gas pressure depending on $M_0$,
we use the kinetic energy flux rather than the total
energy flux to normalize the ``CR energy ratio''.

In Figure 8 we show the CR energy ratio, $\Phi$,
the CR pressure at the shock normalized to the ramp pressure of the upstream
flow in the instantaneous shock frame,
$P_{\rm c,2}/\rho_0 (u_s^{\prime})^2$,
and the ``time-averaged'' injection efficiencies, $\xi$,
for shocks with different Mach numbers when $\epsilon=0.2$ 
(left three panels).
For all Mach numbers the postshock $P_{\rm c,2}$ increases until
a balance between injection/acceleration and advection/diffusion of CRs
is achieved, and then stays at a steady value afterwards.
The time-asymptotic value of the CR pressure becomes, once again,
$P_{\rm c,2}/\rho_0 (u_{s,0}^\prime)^2 \sim 0.56$ and 
$P_{\rm c,2}/\rho_0 (u_s^\prime)^2 \sim 0.8$, for $M_0= 30$ with $\epsilon=0.2$.
We note that the ``undulating'' features in the time evolution of $P_{\rm c,2}$
seem to be numerical artifacts and not real.
Unlike other spatially averaged or integrated quantities, the 
plotted $P_{\rm c,2}$ is sampled exactly at the subshock (\ie from one zone) 
whose properties show small, noisy variations in time.
They seem in particular to correspond to times when the
subshock crosses a regular zone boundary and are most prominent for 
high Mach number models of $M_0=30,$ and 100.
These features can be also seen in the preshock and postshock densities shown
in Figure 7.

The CR energy ratio, $\Phi$, increases with time as CRs are injected
and accelerated, but it asymptotes to a constant value, 
once $P_{\rm c,2}$ has reached a  quasi-steady value.
This results from the approximate ``self-similar'' evolution of the $P_{\rm c}$
spatial distribution.
Time-asymptotic values of $\Phi$ increase with $M_0$ 
and $\Phi\approx 0.5$ for $ M_0=30$ at the terminal time. 

As discussed in \S II.(c)
in the thermal leakage model the injection rate is higher for higher 
subshock Mach number, because the ratio of the thermal peak velocity 
to the injection velocity is smaller. 
This ratio, however, becomes independent of $M_{sub}$ for $M_{sub}>10$,
so the injection rate also shows the same trend.
This Mach number dependence of $\xi$ can be observed in Figure 8,
where the initial value of $\xi$ (at $t=3$) increases with $M_0$, but
it is about the same for $M_0\ge 10$. 
After the initial quick increase, it decreases in time as the 
subshock weakens due to the pre-deceleration in the precursor. 

Once again, in order to explore the dependence of our injection model on the 
parameter $\epsilon$, especially for low Mach shocks,
we also show the results for the $M_0=2$ ($M_s=3$) models 
with $\epsilon=0.2,$ 0.25, 0.3, and 0.4
in the right three panels of Figure 8.
As expected, the injection rate is higher for larger values of $\epsilon$,
so the CR pressure and $\Phi$ are higher. 
In Kang \etal (2002), we made a similar comparison for a wider
range of Mach numbers and found that the dependence on $\epsilon$ is
much weaker for stronger shocks.
In a physical model the parameter, $\epsilon$, should depend 
on the subshock Mach number. 
But it is not well understood how $\epsilon$ should
vary with the subshock Mach number for weak shocks.
For strong shocks, the average injection rate is about $10^{-3}$ with
$\epsilon = 0.2 $, which corresponds to strong wave generation and 
inefficient leakage.  This injection rate is in fact in a good agreement
with what has been observed in the Earth's bow shock (Quest 1988).
For $M_s=3$ shock, the similar injection rate is obtained for 
$0.25 \le \epsilon \le 0.3$, and the CR pressure is about 10-15 \%
of the shock ram pressure, which could be considered substantial. 
Then we conclude that CRs can absorb a significant portion of the shock 
kinetic energy at cosmological shocks, if about $\sim 10^{-3}$ of the particles 
are injected into the CR component regardless of the details of the
injection process.

\section{SUMMARY}

We have calculated CR acceleration at 1D quasi-parallel shocks
by using our CRASH (Cosmic-Ray Amr SHock) code (Kang \etal 2002), 
which incorporates the ``thermal leakage'' injection process to the CR/hydro
code that solves the CR diffusion-convection equation along with CR modified 
gasdynamic equations.
Our simulations are performed in a 1D plane-parallel space in which
shocks are generated by the accretion flow reflecting off the central 
symmetry plane.
For convenience, we considered the accretion velocity of 
$v_{\rm acc}=1500 {\rm km s^{-1}}$ and the magnetic field of 1 microgauss 
as fiducial values for our simulations. 
However, the general conclusions can be applied to similar shock speeds,
because the CR acceleration is controlled mainly by two physical parameters, 
the shock Mach number, $M_s$, and the inverse wave-amplitude parameter, $\epsilon$.
The current simulations are similar to those presented in Kang \etal (2002), 
in which freely propagating shocks with open boundaries were considered,
except that we adopted a reflecting boundary in the downstream region
and our shock velocities, $v_s = 2000 - 2250 {\rm km s^{-1}}$, are lower than 
their value of 3000 kms$^{-1}$. 
Having the reflecting plane expedites the CR acceleration in our
simulations, because CRs are trapped between the shock and the downstream, 
reflecting, boundary, but otherwise the results are mostly similar.  
In particular we conclude:

1) Suprathermal particles can be injected very efficiently
into the CR population via the thermal leakage process, so that typically
a fraction of $10^{-4} - 10^{-3}$ of the particles passed through the shock 
become CRs for $ 0.2 \le \epsilon \le 0.4$. 

2) For a given value of $M_s$, 
the injection efficiency is higher for larger values of $\epsilon$ (\ie weaker waves) 
and so the CR acceleration is more efficient. 
For example, in the shocks with $M_s = 3$, 
the ratio $P_{\rm c,2}/ \rho_0 (u_s^\prime)^2$ becomes 
0.1, 0.16, and 0.24 for $\epsilon=0.25,$ 0.3, and 0.4, respectively.  
But this dependency is weaker for higher Mach numbers, 
so the acceleration efficiency becomes approximately independent of $\epsilon$ 
in the strong shock limits ($M_s \gsim 30$). 

3) For a given value of $\epsilon$, the acceleration efficiency
increases with $M_s$, but it asymptotes to a limiting value for $M_s \gsim 30$.
For example, the model with $\epsilon=0.2$,
the ratio of $P_{\rm c,2}/ \rho_0 (u_s^\prime)^2$ becomes 
0.03, 0.3, 0.5, 0.8, and 0.9, and 
the ratio of $P_{\rm c,2}/P_{\rm g,2}$ approaches to 
0.05, 0.6, 1.5, 4.9, and 13  
for $M_s =$ 3, 6.8, 13.3, 40, and 133, respectively. 

4) In the strong shock limit of $M \ge 30$,
the CR pressure can dominate over the gas pressure and induce a significant
precursor where the preshock flow is decelerated adiabatically. 

5) The CR pressure seems to approach a time asymptotic value when a
balance between acceleration/injection and diffusion/advection processes is
achieved, resulting in an approximate ``self-similar'' flow structure.
This is achieved in a time scale comparable to the acceleration time scales
for the mildly relativistic protons ($p/m_{\rm p}c \sim 10$),
which is much shorter than the cosmological time scale. 

We suggest that the CR acceleration at the cosmic shocks are innate to
collisionless shock formation process and CRs can absorb a significant
fraction of dynamical energy associated with the gravitational collapse during
the formation of large scale structure.
For strong accretion shocks of $M_s > 10$, CRs can absorb most of shock
kinetic energy and the accretion shock speed can be reduced up to 20 \%, 
compared to pure gas dynamic shocks. 
Although the amount of kinetic energy passed through accretion shocks is small, 
since they propagate into the low density intergalactic medium,
they might possibly provide acceleration sites for ultra-high energy cosmic rays
of $E>10^{18}$eV. 
For internal merger shocks of $M_s<3$ the energy transfer to CRs should be less 
than  10-20 \% of the shock kinetic energy at each shock passage, 
with an associated  CR particle fraction of $10^{-3}$. 
Considering that ICM can be shocked repeatedly, however, the CRs generated by 
merger shocks could be sufficient to explain the observed non-thermal signatures
from ICM of galaxy clusters in radio, EUV and X-ray. 
This implies that the current understandings of cosmological hydrodynamic
simulations could be modified by inclusion of this process 
at a quantitative level of order several tens of percent.

\acknowledgments{
This work was supported by Korea Research Foundation Grant 
(KRF-2001-041-D00270).
TWJ is supported by NSF grant AST00-71167, by NASA grant NAG5-10774
and by the University of Minnesota Supercomputing Institute.
}

\end{document}